\documentclass[12pt]{iopart}
\usepackage{graphicx}

\begin{document}

\title{Combined magnetic and chemical patterning for neural architectures}

\author{C Delacour$^{1,}$$^{2}$, G Bugnicourt$^{1,}$$^{2}$, N M Dempsey$^{1,}$$^{2}$, F Dumas-Bouchiat$^{1,}$$^{2}$, C Villard$^{1,}$$^{2}$}
\address{1. Univ. Grenoble Alpes, Inst. NEEL, F-38042 Grenoble, France}
\address{2. CNRS, Inst. NEEL, F-38042 Grenoble, France}
\ead{cecile.delacour@neel.cnrs.fr}
\begin{abstract}
In-vitro investigation of neural architectures requires cell positioning. For that purpose, micro-magnets have been developed on silicon substrates and combined with chemical patterning to attract cells on adhesive sites and keep them there during incubation. We have shown that the use of micro-magnets allows to achieve a high filling factor ($\sim$ 90 $\%$) of defined adhesive sites in neural networks and prevents migration of cells during growth. This approach has great potential for neural interfacing by providing accurate and time-stable coupling with integrated nanodevices.

\end{abstract}

\submitto{\JPD}
\maketitle

\section{Introduction}
Chemical micro-patterning is a versatile technique to control cell positioning and cell spreading in vitro. In the case of neurons, the geometry of micro-patterns must combine large (i.e. in the range 20-$80\mu m$ in diameter) spots dedicated to cellular bodies (soma), and micron-sized stripes that ensure neuronal connections \cite{wyart2002constrained, Roth2}. Although both the architecture and the polarity of these networks are now well mastered at the single cell level \cite{Roth2}, a remaining challenge in the field of neuronal network design is to establish a robust method to fill all soma sites.\\
Seeding cells on top of micro-patterns relies on the sedimentation of a cell suspension. The inherent low yield in the filling of adhesive sites related to the stochastic nature of the sedimentation process \cite{Lutoff2011} could in principle be increased by adjusting the initial concentration of the cell suspension. However, in the specific case of neurons, this would lead cells to land and to stick on the thin adhesive paths that will further guide neuronal branches (neurites), therefore impairing the initial network design. To circumvent this problem, an active cell positioning technique is required. Besides, the mechanical tensions involved in the establishment of neuronal connections could destabilize the initial cellular architecture through the forces exerted by the neurites on soma \cite{Roth2}. Therefore, there is a need for a trapping methodology that would keep soma on their adhesive sites throughout the incubation period. Lastly, in the perspective of neuronal recording, an active positioning technique must be compatible with the implementation of a sensor array.\\
In the past few years, several methods have thus been developed for active cell entrapment \cite{Nilsson} such as dielectrophoresis \cite{wang2013adjustable}, optical \cite{padgett2011holographic} or magnetic tweezers  \cite{winkleman2004magnetic, Kilgus}. They have been used to trap both single and multiple cells at defined positions on almost any kind of substrate. However, these methods cannot be used throughout the maturation time of neurons. They should be combined with surface traps such as holes \cite{wang2012positioning} or cages \cite{zeck2001noninvasive} to immobilize cells, and these topographies are hardly compatible with the complex surface of a sensor array. 
Long lasting traps and neurite guidance have been achieved with microfluidic systems \cite{lin2013microfluidic} but the shear stress encountered even under moderate flow disturbs neuronal growth and remains a major obstacle to its wide use \cite{wang2008microfluidics}. Micro-patterned soft and hard magnetic films have been used to trap cells functionalized with magnetic nanoparticles (for a recent review see \cite{chen2014multiscale}). Such traps may be exploited for long lasting and multi-site trapping of neurons, and if combined with conventional silicon electronics, they could be used in neuronal network studies. Note that while soft magnetic elements need to be magnetized by an external magnetic field source, hard magnetic elements do not, rendering them more compact and well suited to time stable trapping. What is more, the stray magnetic fields produced by hard micro-magnets are confined to the region of interest, limiting the risk of potential interference with other components of the device. We have previously reported on the fabrication of arrays of hard micro-magnets by patterning of hard magnetic films through the use of pre-patterned substrates \cite{Walther} or Thermo-Magnetic-Patterning \cite{Dumas}, or the Micro-Magnetic-Imprinting of hard magnetic particles \cite{dempsey2014micro}. Such arrays have been used for the trapping of various types of magnetically functionalized cells \cite{zanini2012micromagnet, osman2012monitoring, pivetal2014selective, dempsey2014micro}, the diamagnetic trapping of non-functionalized Jurkat cells in the presence of a paramagnetic contrast agent \cite{kauffmann2011diamagnetically} and the networking of non-functionalized stem cells in culture medium \cite{zablotskii2013life}.\\In this work, we report on the fabrication of arrays of isolated micro-magnets by the lift-off process, to allow contactless, multi-site, time-stable entrapment of cells on silicon substrates. We combine this magnetic patterning process with a chemical micro-patterning process designed to direct neuronal growth and control cell polarity \cite{Roth2, Roth1}. Such an approach should provide both confinement and neurite guidance above silicon nanosensors while allowing the use conventional seeding and culturing protocols.\\
Our method aims to use the magnetic interaction between micrometer sized magnets and dissociated cells tagged with micro-sized beads containing typically 30 vol \% Fe-oxide superparamagnetic nanoparticles (spm-NPs). When a cell comes close enough to a micro-magnet, it gets attracted and trapped. The adhesion site is also coated with adhesion-promoting molecules that will further guide elongation of neuronal branches on defined pathways \cite{wyart2002constrained, Roth2}. The force exerted by a micro-magnet on a bead is given by:
\begin{equation}
\label{eq:force}
\mathbf{F} = \mu_0.V.(\mathbf{M}.\mathbf{\nabla})\mathbf{H}
\end{equation}
where $\mu_0$ is the magnetic permeability constant, V the total volume of spm-NPs, \textbf{M} the magnetization of a given spm-NP which depends on \textbf{H}, the field produced by the micro-magnet. Therefore, the strength of the magnetic force increases with both the magnetic field and its gradient.
Reducing the size of a magnetic field source while maintaining its field strength leads to an increase in the field gradient it produces. A reduction of all dimensions by a factor k leads to an increase of the magnetic field gradient, and thus the volumic force it exerts on a given object, by the same factor k. Micro-patterning of Nd-Fe-B hard magnetic films produces magnetic field gradients as high as $10^{5}$-$10^{6}$ T/m  \cite{NdFeB}. When wanting to integrate micro-magnets into a given device, one needs to consider the compatibility of the processing temperature of the magnetic material with other device components. The relatively high processing temperature required to crystalize the high anisotropy $Nd_{2}Fe_{14}B$ phase (T $\ge$ 500\char23C) \cite{Dempsey}, precludes its use in certain studies. SmCo hard magnetic films can be processed at lower temperatures (300-400\char23C in the directly crystallized state) \cite{walther2008structural}; $\geq$ 450\char23C for films deposited in the amorphous state and crystallized in a post-deposition step, rendering their fabrication compatible with conventional micro-technology and front-end integration on silicon chips. Here we report on the micro-patterning of SmCo films and the subsequent chemical patterning of the host Si wafer for the controlled seeding and subsequent growth of neurons.

\section{Experiment}
\subsection{Micro-magnet fabrication and characterization}
\begin{figure}[h!]
\centering
\includegraphics[scale=0.4]{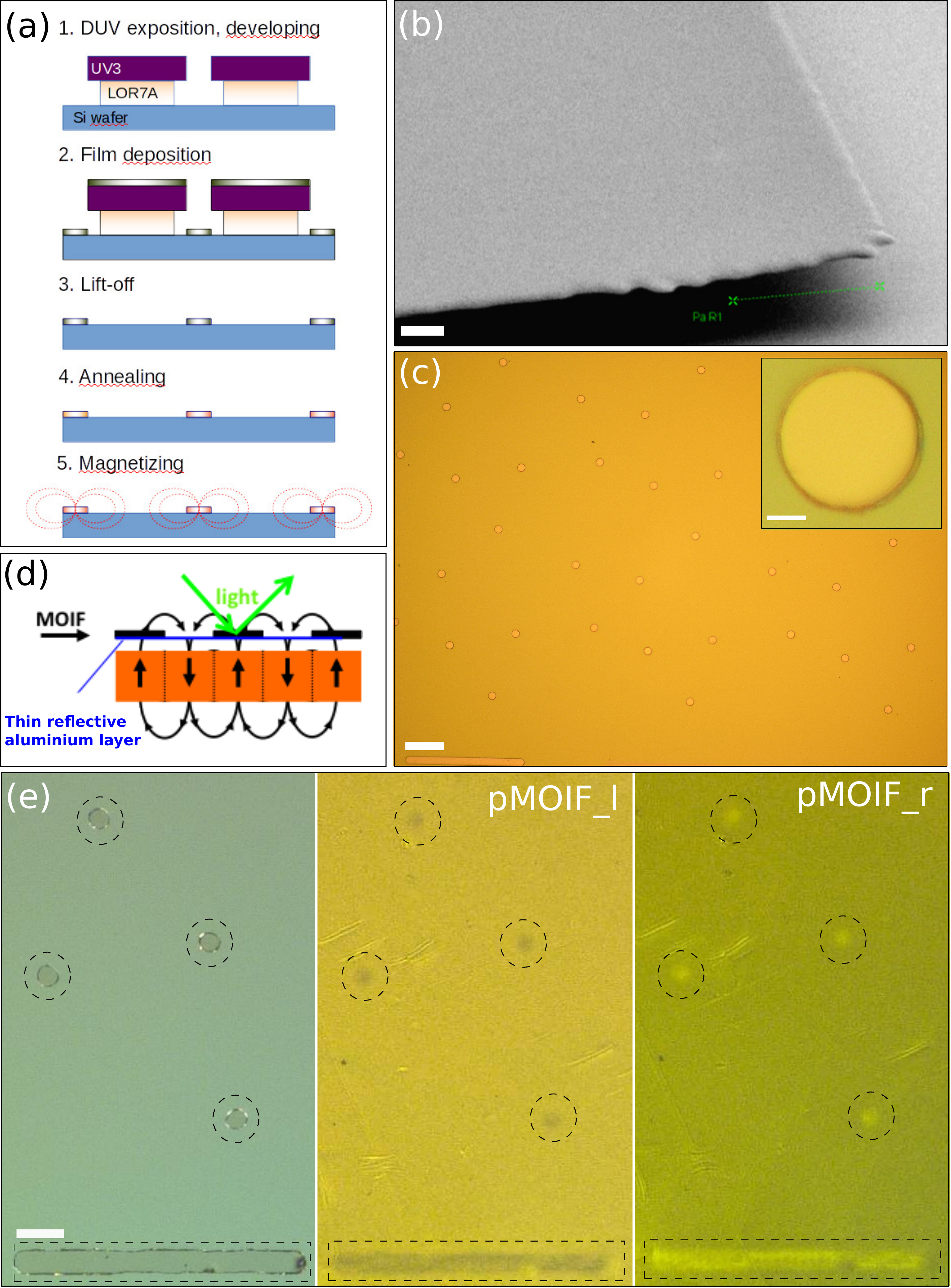}
\caption{\label{fig:proto}(a) Flow chart of micro-magnet fabrication. (b) SEM micrography showing the large undercut obtained with the bi-layer resist (UV3/LOR). Scale bar is 1.5 $\mu$m. (c) SmCo micro-disk array ($\phi$ = 20 $\mu$m). Scale bar 100 $\mu$m. Inset: optical micrography of a micro-disk. Scale bar 5 $\mu$m. (d) Principle of Magneto-Optical Imaging Film (MOIF). (e) From left to right: direct optical and magneto-optical micrographies of a planar MOIF placed on top of the micro-magnets. The analyser was rotated left (l) then right (r). Magnetic patterns are outlined for sake of clarity. Scale bar 50  $\mu$m.}
\end{figure}

Figure~\ref{fig:proto}.a illustrates the micro-magnet fabrication process. Firstly, silicon substrates are cleaned and spin-coated with a bi-layer resist (LOR/UV3). The resist is removed in solvent after DUV exposure producing a large undercut (fig.~\ref{fig:proto}.b). Secondly, Cr/SmCo/Cr films, here after simply referred to as "SmCo" films (the Cr buffer and capping layers serve to prevent diffusion into the Si substrate and oxidation, respectively) are deposited by triode-sputtering \cite{walther2008structural}. Film lift-off (3) is obtained by removing the bilayer resist in an acetone bath. Figure~\ref{fig:proto}.c shows an example of a 400 nm thick SmCo micro-disk array obtained with a bottom LOR-resist layer of 700 nm thickness. Lift-off produces well-formed micrometer sized magnets with relatively sharp edges (Inset figure~\ref{fig:proto}.c) while avoiding contamination or mechanical damage that could result when etching or polishing the magnetic film. The last fabrication steps are (4) annealing at 600\char23C (1h) of SmCo (note that this temperature can in principle be reduced at 450\char23C, while maintaining hard magnetic properties), and (5) magnetization of the micro-magnets in an out-of-plane oriented magnetic field (4T). \\The hard magnetic nature of the SmCo micro-disks is revealed using magneto-optical imaging with the aid of a planar Magneto-Optic Imaging Film (MOIF). The Magneto-Optical Imaging Film (MOIF) used here is a soft magnetic bismuth-substituted ferrite-garnet film covered with a reflective thin aluminium layer. The localized stray magnetic fields produced by the hard magnetic SmCo disks generate characteristic domain patterns in the MOIF. These magnetic domains are revealed in a magneto-optical microscope through the Faraday effect, which causes a rotation of the plane of polarization of the incident light (fig.~\ref{fig:proto}.d). Note that the use of a capping layer to prevent oxidation of the SmCo micro-disks precludes direct Kerr imaging of the micro-disks themselves. A rotation of the analyser through the angle of extinction induces a change in contrast from bright to dark, confirming the magnetic nature of the observed contrast (fig.~\ref{fig:proto}.e). The MOIF-based method, reviewed by Grechishkin et al \cite{grechishkin2008magnetic} offers an attractive combination of simplicity, sensitivity and spatial resolution at high acquisition rates.

\subsection{Surface patterning}
We have combined the use of micro-magnets with chemical patterning to build defined neural networks. Micro-magnets are first encapsulated in a biocompatible layer of parylene-C (30 nm) deposited by cold CVD in a LabTop Model 3000 from Para-Tech  \cite{delivopoulos2011controlled}. Then, we aligned the poly-$_{L}$-lysine (PLL) pattern on the micro-magnet array with conventional DUV-optical lithography and lift-off. The PLL patterns promote neurite elongation along defined pathways after cell trapping. The purpose of their specific design, combining curvatures and non-circular soma adhesive spots (fig.~\ref{fig:Neuron_aimant_5}) is explained in \cite{Roth2}. 

\subsection{Cell culture}
Primary hippocampal neurons were prepared from E18 mouse embryos \cite{kaech2007culturing}, mixed with PLL-coated magnetic beads and plated at the concentration of 2.$10^{4}$ cell/$cm^{2}$. No internalization was observed during time-lapse experiments (i.e. video recording over days) of micro-bead attachment. This is not surprising considering that the size of the beads used in this study ($\sim$ 3  $\mu$m) is almost half the size of the soma. We estimated that 40 $\%$ of plated cells are tagged with beads by counting the number of cells decorated with at least one bead. This estimation was performed by taking a small volume of the working mixture of PLL-coated beads (incubated over night at room temperature using a PLL concentration of 1 mg/ml) and cells (10 min incubation with the PLL-tagged beads in a 10:1 bead:cell proportion) and letting the mixture sediment on a glass coverslip. After 3 days, neurons were fixed with 3.7 $\%$-formaldehyde in phosphate-buffered saline (PBS, pH 7.2) for further investigations of micro-magnet trapping efficiency and neurite guidance on PLL-patterns. 
\section{Results and discussion}
\subsection{Neuron positioning}
\begin{figure}[h]
\centering
\includegraphics[scale=0.5]{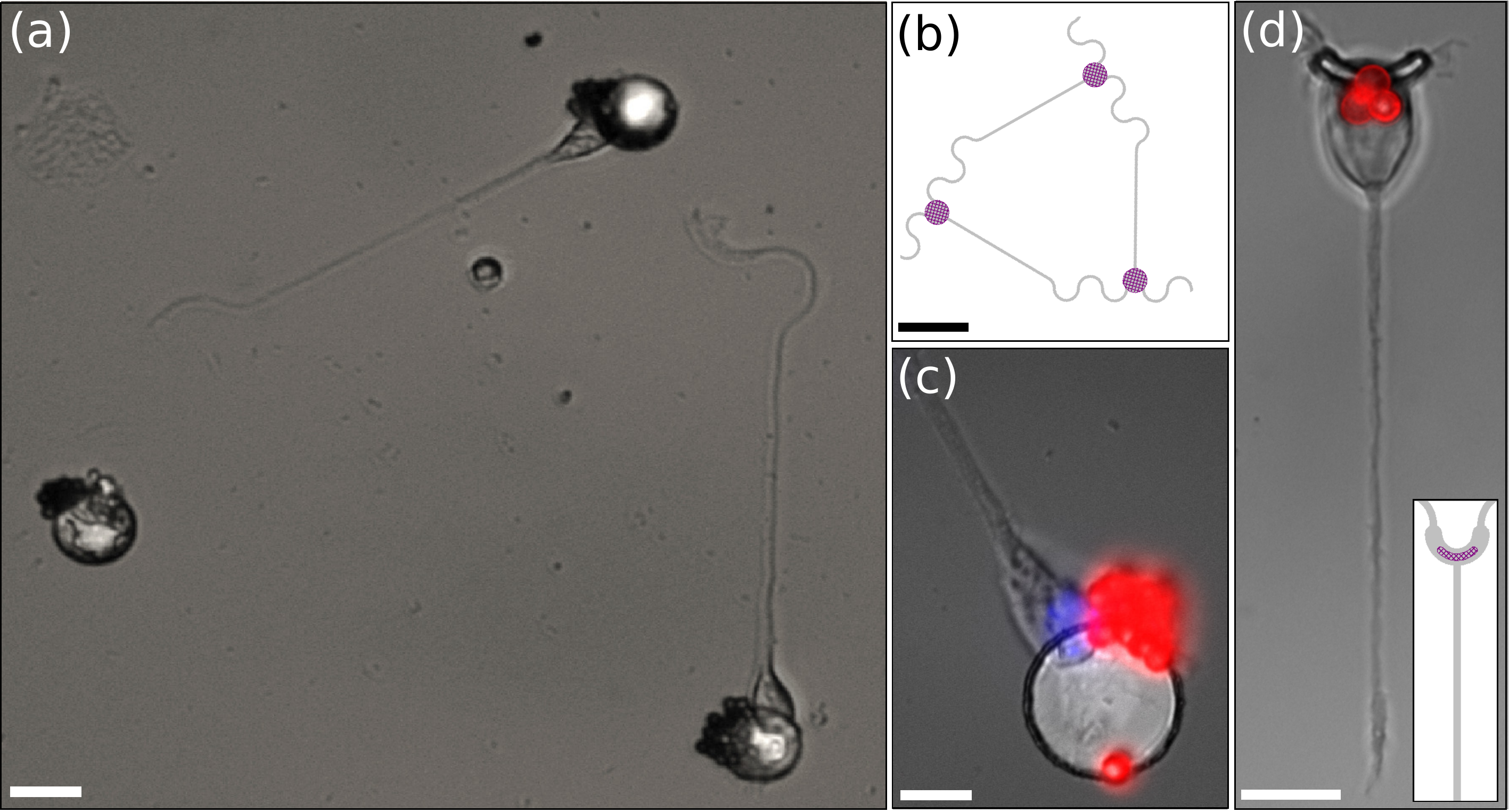}
\caption{\label{fig:Neuron_aimant_5}(a) Optical micrography of neurons (3-DIV) trapped on SmCo hard micro-magnets and growing on the Poly-$_{L}$-lysine (PLL) adhesive pattern. Scale bar 20 $\mu$m. (b) Superimposed patterns of PLL (gray) and  micro-magnets (hatched purple). Scale bar 60 $\mu$m. (c) Superimposed nuclei staining (Hoechst, blue) and bead fluorescence (red). Scale bar 10 $\mu$m. (d) Neuron trapped on a boomerang shaped micro-magnet ($\phi$ = 2 $\mu$m) and growing on a PLL straight pattern. Inset: schematic of PLL and micro-magnet patterns. Scale bar 10 $\mu$m.}
\end{figure}

 An example of magnetic cell trapping is shown in figure~\ref{fig:Neuron_aimant_5}.a  for a triangular neural network (fig.~\ref{fig:Neuron_aimant_5}.b). Somas are well placed on the SmCo hard micro-magnets. Staining of nuclei (blue) and beads (red) outline their position on the micro-magnets (fig.~\ref{fig:Neuron_aimant_5}.c). They are attracted to the edges of the micro-magnets, where the stray magnetic field and magnetic field gradient are the highest (eq~\ref{eq:force}). In addition, a long neurite, presumably the axon \cite{Roth2}, has emerged from two of them and elongated along the PLL stripes of the pattern. Figure~\ref{fig:Neuron_aimant_5}.d shows the trapping of a neuron on a boomerang shaped  micro-magnet (2 $\mu$m wide) designed to control the sub-cellular organization of the soma \cite{Roth2}. 

\subsection{Trap efficiency}
\begin{figure}[h]
\centering
\includegraphics[scale=0.5]{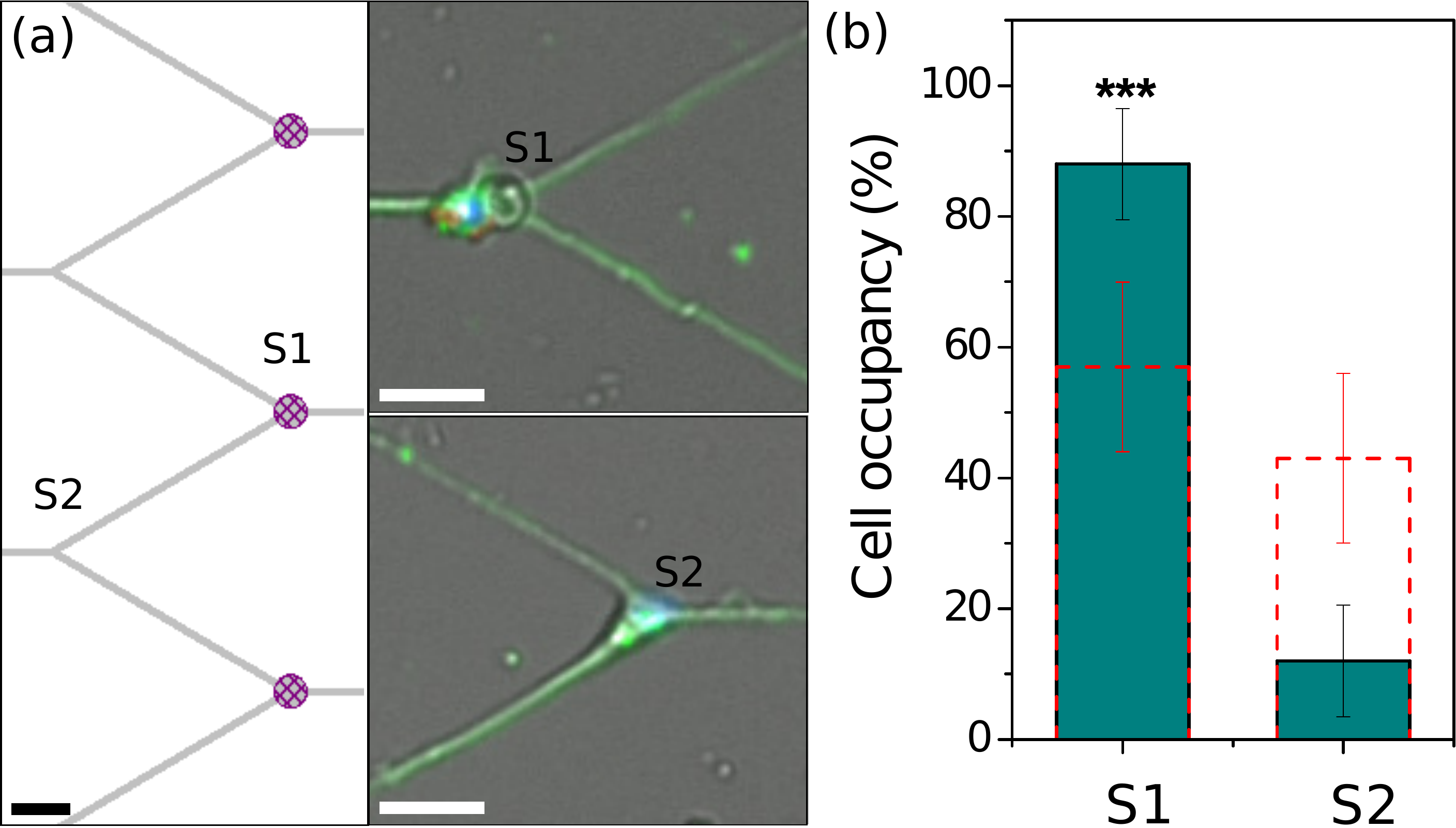}
\caption{\label{fig:placement_stat} (a) Patterns used for statistical study. Right : Schematic of the PLL pattern (gray) coinciding with micro-magnets (hatched purple) (S1)  and the PLL crossings (S2). Left : Superimposed nuclei (Hoechst, blue), microtubulin (green) staining and bead fluorescence (red) for 3-DIV neurons seeded above the combined  SmCo micro-magnet array and PLL micro-patterns. Scale bars 40 $\mu$m. (b) Percentages of cell occupancy per sites S1 and S2, without magnetizing micro-disks (dashed line), with magnetized micro-disks (green columns). Error bars denote $95\%$ confidence interval. *** indicates p $<$ 0.001.}

\end{figure}

Random seeding of neurons on PLL patterns could increase the filling ratio of the networks. To estimate the relative contributions of the magnetic and chemical traps, we have compared the number of cells on PLL coated micro-magnets (S1) and on crossings covered by PLL only (S2) as described in figures~\ref{fig:placement_stat}.a. Firstly, we performed the experiment on a control sample to probe the effect of chemical (PLL) trapping only. For this test, we stopped the sample fabrication at step 4 (fig.~\ref{fig:proto}.a), i.e. no external magnetic field was applied to magnetize the SmCo micro-disks. In the virgin state, the domain size in the SmCo micro-disks is of the order of 600 nm, as estimated by Magnetic Force Microscopy (data not shown). As both the field and the field gradient produced by a magnetic domain drop off quickly with distance from the domain, the magnetophoretic forces produced by non-magnetized micro-disks, compared to magnetized micro-disks ($\phi$ = 20 $\mu$m), are considered to play a negligible role in attracting neurons to the target sites. For such non-magnetized samples, 57 $\%$ of neurons are trapped on the micro-disks and 43 $\%$ on the crossings (n = 108 cells). The probability to catch a cell on a micro-disk is slightly higher because of the larger adhesive surface and topographic effects. Then, we repeated the experiment after magnetization of the SmCo micro-disks. The resulting hard micro-magnets are covered with the same PLL pattern. We counted 49 tagged cells trapped on micro-magnets (S1) but only 7 on PLL crossings (S2): 88 $\%$ of trapped cells are magnetically attracted (n = 56 cells). This result is significantly higher than the 57 $\%$ obtained on the control sample (p $<$ 0.001, z test)(fig.~\ref{fig:placement_stat}.b). \\
However, this conclusion has to be further refined. Among the counted cells some could result from independent events: (1) trapping of free beads on the micro-magnet followed by (2) bead free cell trapping on the PLL coated micro-disk. We have beads without cells on 37 $\%$ of micro-magnets ($P_{1}$) and 57 $\%$ of cells on the PLL-coated micro-magnet ($P_{2}$) (control sample). Thus we estimate that 21 $\%$ ($P_{1}.P_{2}$) of the 49 counted cells may result from these independent events. Therefore 85 $\%$ of cells are well-positioned magnetically, instead of 88 $\%$ (p $<$ 0.01, z test). The final result remains still significant compared to the control sample and demonstrates the great potential of such micro-magnets. The use of micro-magnets increases by nearly 4 times  (0.85 $^{3}$ vs 0.57 $^{3}$) the probability to get a fulfilled 3-neuron network. We expect that a preliminary sorting of tagged and non-tagged cells and/or  microfluidic guiding of tagged cells close to the target sites would enhance magnetically induced site filling. In addition, filtering of unattached beads by inclusion of a centrifugation step would reduce the stochastic filling of adhesive patterns by the beads, providing controlled and uniform conditions for cell attachment on soma adhesive sites.

\section{Conclusion}
SmCo hard micro-magnets have been fabricated on silicon substrates with conventional optical lithography and the lift-off technique. We have combined micro-magnets and chemical patterning to position neurons to form networks of controlled architectures. This method has the potential to attract and keep somas at their initial place throughout the incubation time. A microfluidic channel may be used to increase the efficiency of magnetic trapping, by guiding neurons close to the magnetic traps. Such an approach is compatible with a wide range of substrates such as silicon. Lastly, this magnetic assisted cell positioning technique provides an efficient way to couple organized cell networking and integrated nanosensors on silicon with the required time-stability for long term neural interfacing.

\ack We thank D. Givord for helpful discussions concerning neuro-magnet coupling, E. Andre and T. Crozes for fabrication support, J. Brocard and N. Collomb who provided excellent cell support.

\section*{References}
\bibliography{NeuroMagnet}
\bibliographystyle{unsrt} 

\newpage

\end{document}